\documentclass[useAMS,usenatbib]{mn2e}

\usepackage{graphicx}
\usepackage{epstopdf}
\usepackage[footnotesize]{subfigure}
\usepackage{amsmath}
\usepackage{float}

\def\gs{\mathrel{\raise0.35ex\hbox{$\scriptstyle >$}\kern-0.6em
\lower0.40ex\hbox{{$\scriptstyle \sim$}}}}
\def\ls{\mathrel{\raise0.35ex\hbox{$\scriptstyle <$}\kern-0.6em
\lower0.40ex\hbox{{$\scriptstyle \sim$}}}}
\def\ls{\mathrel{\hbox{\rlap{\hbox{\lower4pt\hbox{$\sim$}}}\hbox{$<$}}}}
\def\gs{\mathrel{\hbox{\rlap{\hbox{\lower4pt\hbox{$\sim$}}}\hbox{$>$}}}}

\def\mnras {{\sc MNRAS}}
\def\apj {ApJ}
\def\apjs {ApJS}
\def\apjl {ApJL}
\def\aj {AJ}
\def\aaps {A\&AS}

\def\aap {A\&A}

\newcommand\gtsim{\mathrel{\lower0.6ex\hbox{$\buildrel {\textstyle >}
  \over {\scriptstyle \sim}$}}}
\newcommand\ltsim{\mathrel{\lower0.6ex\hbox{$\buildrel {\textstyle <}
  \over {\scriptstyle \sim}$}}}

\title[Post merger properties of A1664 cluster galaxies]
      {Photometric studies of Abell 1664: The subtle effect a minor merger has on cluster galaxies}
\author[D.\, Kleiner et al.]
       {Dane Kleiner,$^{\! 1, 2, 3}$\thanks{E-mail: dane.kleiner@monash.edu}
	Kevin A.\ Pimbblet,$^{\! 1, 2, 4,5}$
	Matt S. Owers,$^{\! 6}$
	D. Heath Jones$^{\! 1, 2}$ 
	\newauthor and Andrew P. Stephenson$^{\! 7}$	
        \vspace*{1mm}\\
        $^1$ School of Physics, Monash University, Melbourne, Victoria 3800, Australia\\
        $^2$ Monash Centre for Astrophysics, Monash University, Melbourne, Victoria 3800, Australia\\
        $^3$ CSIRO Astronomy \& Space Science, Australia Telescope National Facility, P.O. Box 76, Epping, NSW 1710, Australia\\
        $^4$ Department of Physics, University of Oxford, Denys Wilkinson Building, Keble Road, Oxford OX1 3RH, UK\\
        $^5$ Department of Physics and Mathematics, University of Hull, Cottingham Road, Kingston-upon-Hull, HU6 7RX\\
        $^6$ Australian Astronomical Observatory, P.O. Box 915, North Ryde, NSW 1670, Australia\\
        $^7$ School of Mathematics and Physics, University of Queensland, QLD 4072, Australia\\
}

\date{Accepted 2014 January 16.  Received 2014 January 15; in original form 2013 December 2}

\pagerange{000--000}

\begin{document}

\maketitle

\begin{abstract}
A combination of $BRI$ photometry and archival Chandra X-ray data have been used to analyse the effects a minor merger has on the galaxy population of A1664. We utilise adaptive smoothing techniques in the 2D spatial distribution of cluster galaxies to reveal substructure $\sim$ 800 kpc South of the cluster core. We identify this substructure as most likely the remnant core of a merging group which has passed pericentre and responsible for triggering a cold front in the cluster core. We define two samples to represent two different environments within A1664 in accordance with the location of the substructure. We apply a morphological analysis using CAS, M$_{20}$ and Gini to these samples to deduce if there has been any significant effect on the cluster galaxies due to this interaction. We find there are more asymmetric galaxies found in the inner sample (at the 3.7$\sigma$ level) which is likely due to galaxy-galaxy interactions as the merging group passed through core passage. No other differences were found between the inner and outer cluster in our morphological analysis, which we attribute to the limited resolution of our imagery. The colour profiles of the galaxies are found to be consistent with the morphology-density relation suggesting there is no unique environmental effect in A1664 that has enhanced galaxy transformations. This study favours the star formation of cluster galaxies being quenched well before it is able to interact with the merging group and demonstrates that a minor cluster merger has little effect on the observable parameters of cluster galaxies such as morphology and colour. 
\end{abstract}


\section{Introduction}
In a Universe where structure forms through hierarchical growth, galaxy clusters are the largest  structures which are virialised at the current epoch. Individual clusters can offer the opportunity to study the properties of hundreds of galaxies in the densest environment in the Universe. It has been well documented that the cores of low redshift clusters are heavily populated by red early type galaxies and display significantly different properties to galaxies found in less dense environments (Dressler 1980; Balogh et al.~2004; Poggianti et al.~2006). The rich cluster environment hosts a number of interactions (both galaxy-galaxy and cluster-galaxy) that offer the potential to explain the difference in observed galaxy properties between galaxies found in over- and under-dense environments. For example galaxies that are close enough to gravitationally interact can destroy spiral discs and induce an episode of star formation (Toomre \& Toomre 1972). Strong tidal interactions between galaxies can strip gas from the interacting galaxies in long tidal tails (Barnes \& Hernquist 1991) and dwarf galaxies that interact with much more massive galaxies can be gravitationally shocked, removing up to 50\% of its gas content (Moore et al.~1996). As galaxies travel through the ICM, the impact with the hot intra-cluster gas can strip the gas contained in the disc (Gunn \& Gott 1972, Nulsen 1982) or remove the reserve gas supply in the halo preventing the reservoir of halo gas to condense in the disc and form new stars (Larson et al.~1980).

One important result in this area is that cluster galaxies at large cluster-centric radii are found to have a observable properties such as colour, morphology and star formation rate (SFR) that have already been truncated (Lewis et al.~2002; Gomez et al.~2003; McGee et al.~2009; Wolf et al.~2009; Lu et al.~2012; Rasmussen et al.~2012). This conclusion shifted the focus away from transformations in the dense cluster environments being the only reason that explains the difference of galaxy properties in different environments. Less dense environments such as galaxy groups in the outskirts of clusters (Rines et al.~2005; Mahajan et al.~2012; Verdugo et al.~2012) and filaments of galaxies (Porter \& Raychaudhury 2007; Edwards et al.~2010; Biviano et al.~2011) are now considered favourable sites for first-time tidal interactions and are thought to play a vital role in explaining the difference in galaxy population. Through these interactions, the star formation fuel of a galaxy can be used up in a starburst (i.e. Porter et al.~2008) or through quenching (i.e. Lu et al.~2012) at these sites. Galaxies that have experienced these transformation have been ``preprocessed" prior to accreting to the cluster potential. 

There exists alternative evidence to suggest that preprocessing is not the main cause for the difference in galaxy populations. Von der Linden et al.~(2010) and Haines et al.~ (2013) show that a galaxy's SFR can be quenched slowly within the cluster and the morphology-density relation is weak or non-existent beyond the clusters virial radius. Berrier et al.~(2009) show that preprocessing is not efficient enough to explain the full observed difference between cluster and non-cluster galaxies where physical cluster mechanisms must play a significant role in explaining the high fraction of red early types in dense environments. It is not well known how efficient preprocessing is and what fractions of galaxies are transformed through cluster mechanisms.

An additional ingredient in this discussion about cluster galaxy evolution is the merger of galaxy clusters themselves. It is a rare occurrence for two clusters of equal mass to merge. However, when a major merger occurs the result is a violent reassembly of the cluster (Markevitch et al.~1999). Minor mergers (i.e. a group merging with a cluster) are more `gentle' and common. A significant fraction of a clusters final galaxy population is therefore accreted through groups merging with the cluster (McGee et al.~2009). Compared to more dynamically relaxed clusters, a merging cluster provides a more hostile environment for the resident galaxies. Previous studies (e.g. Caldwell \& Rose, 1997; Poggianti et al.~2004; Hwang \& Lee 2009; Owers et al.~2011;2012) have shown that merging clusters have the ability to transform galaxies much more efficiently than a cluster in relative isolation and calm environs. Numerical simulations have shown that group-cluster mergers can cause a secondary starburst in cluster galaxies (Bekki 1999) and while the merging group is traveling through pericentre, galaxy-galaxy collisions and ram pressure stripping are enhanced (Vijayaraghavan \& Ricker 2013). As the nature of cluster mergers is complex, the full effect that this reconfiguration has on individual galaxies is yet to be fully understood. 

\begin{figure*}
\centering
{\includegraphics[scale=0.95]{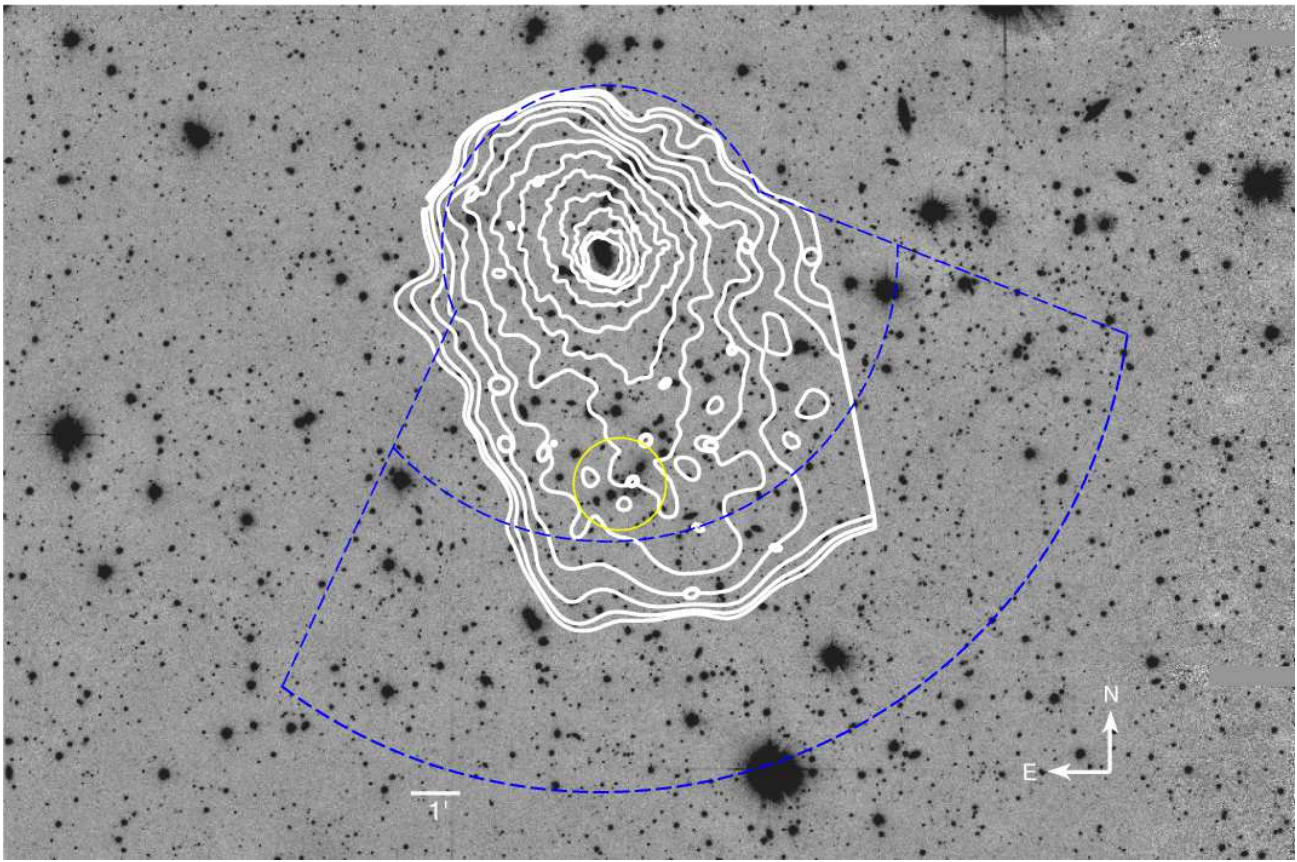}}
  \caption{\small{Coadded $I$-band image of A1664 showing gaussian smoothed X-ray surface brightness contours from an archival Chandra X-ray image in the 0.5 - 7 keV range. The X-ray emission traces the hot intra-cluster gas and shows an elongated morphology which extends to the South-West. The blue dashed regions outline the limits used to define the inner and outer samples in the photometric analysis. The yellow circle encloses a group of galaxies which we identify as the most likely merging group responsible for the elongated X-ray emission.}}
  \label{fig:img}
\end{figure*}

In this paper we analyse Abell 1664 (hereafter A1664), a nearby ($z$ = 0.128; Allen et al.~1995) post-merger galaxy cluster that offers a rare opportunity to study galaxy evolution afforded by its merger status. It has an Abell richness of 2 (Abell et al.~1989) with a velocity dispersion of  $\sigma_{v}$ = 1069$_{-62}^{+75}$ km s$^{-1}$ (Pimbblet et al.~ 2006). A1664 is very bright in the 0.1 -- 2.4 keV band with an X-ray luminosity of 4.1 $\times$ 10$^{44}$ erg s$^{-1}$ (Allen et al.~1992) where the X-ray morphology is elongated in the South (Figure \ref{fig:img}). As the X-ray emission traces the hot intra-cluster gas, the asymmetry in the morphology reveals that A1664 is not a relaxed cluster and the extension in the gas is likely to have once belonged to the merging group. When a cluster undergoes a minor merger, the gravitational disturbance can trigger gas ``sloshing" in the cluster core. If the cluster hosts a dense cool core, this sloshing can generate contact discontinuities at the interface of the displaced cool core gas and the hotter, less dense gas at larger radius (Ascasibar \& Markevitch 2006). These contact discontinuities, otherwise known as cold fronts can be used to identify recent cluster mergers (Owers et al.~2009). Kirkpatrick et al.~(2009) show there to be a sloshing cold front in the core of A1664. They further show the brightest cluster galaxy (BCG) with an unusually high SFR can be explained by a flow of cold gas being deposited on the BCG.

In this case study, our goal is to determine how different environments in a post merger cluster have affected the properties of the galaxies in A1664. Section 2 describes the data analysis and sample selection of the $BRI$ photometric data and \emph{Chandra} image. In  Section 3, we present the analysis of the galaxy properties in different environments. In Section 4, we present the discussion and interpretation of our results on the effect a merging group has on cluster galaxies. Lastly, in Section 5 we summarise our findings.

Throughout this paper, we use a $\Lambda$CDM cosmology where $H_{0}$ = 70 km s$^{-1}$ Mpc$^{-1}$, $\Omega_{m}$ = 0.3 and $\Omega_{\Lambda}$ = 0.7. At this redshift of the cluster ($z$ = 0.128), the assumed cosmology gives 1$\arcsec$ = 2.287 kpc.

\section{Observations and Data Reduction}

\subsection{Optical data}
Our observations of A1664 were made in the $I$-band at the Australian National University 40$^{\prime\prime}$ telescope at Siding Spring Observatory. At the redshift of this cluster ($z =$ 0.128), this bandpass is sensitive to the old stellar populations and dusty galaxies, making it ideal for observing cluster galaxies and their properties. The observations were taken using the wide field imager (WFI). WFI is a mosaic imaging camera that consists of eight 2k $\times$ 4k pixel charged coupled device (CCD) arranged to give a final image that is 8k $\times$ 8k pixel. The pixel size of WFI is approximately 0.376$\arcsec$ $\times$ 0.376$\arcsec$ near the optical axis of the 40$^{\prime\prime}$ which gives an impressive field of view of 52 square arcmin (1.2 degree on the diagonal)\footnote{Given the wide field of view afforded by WFI, the pixel scale varies by up to 9 per cent from the optical axis to the edge of the CCD array.}. The eight individual CCDs are non-linear by $\sim 5$ per cent over their dynamic range. 

Reduction is performed in the standard manner using the {\sc mscred} package in {\sc IRAF}. Briefly, the steps involved include a bias subtraction, a linearity correction followed by flat-fielding of the data using twilight flats. Since WFI is operated at a relatively warm temperature ($\approx 183$ Kelvin), a dark subtraction is also performed. After processing, our science images are flat to better than 1 per cent of the sky background across the entire mosaic.

A world coordinate system (WCS) was created from scratch by matching an average of 5 stellar positions to US Naval Observatory (USNO; Monet et al.\ 2003) stellar catalogue per WFI CCD. The WCS were refined by matching up bright, non-saturated stars using the {\sc msctpeak} task and eliminating stars with high proper motions. Our residual fits for the WCS are typically much better than 0.6 arcsec; rms $\approx 0.2$ arcsec. By employing this WCS,  our multi-extension fits file imaging is then turned in single fits file imaging with {\sc mscimage} which corrects for the radial distortion present in the raw data. After the elimination of bad pixels using the {\sc fixpix} routine, we register our observations with {\sc msczero} and combine them utilising {\sc mscstack}

The final science images are aligned in each passband and trimmed to the same size. Cataloguing of the cluster is performed with {\sc SExtractor} (Bertin \& Arnouts 1996) to permit the detection and photometry of objects in the $I$-band. The final photometry is calibrated from at least 15 sets of standard star observations selected from Landolt (1992) scattered throughout each night. Despite some unstable seeing conditions for our observing run each night was photometric with a seeing of $\ltsim 2$ arcsec where the seeing was measured as the median from the point spread function of the stars as measured with {\sc SExtractor} (Bertin \& Arnouts 1996). Zeropoint and colour term equations used to calibrate our observations yield a final average error of $\sim 0.05$ mags. We consider this calibration error to be the dominant source of photometric error.

To create a coadded $I$-band image, a bad pixel mask was created and assigned to each individual image such that no additional noise was added in the stacking process. We were selective with images used in the stacking processes such that only the images with a seeing of less than 2.5 arcseconds were combined using the \emph{Swarp} package (Bertin et al.~2002). The photometry from a Hubble space telescope (HST) $I$-band image that contained overlapping objects in our field was used to calibrate the RA, DEC and magnitudes. Between $16 < I_{HST} < 20$ the calibrated RMS is 0.15 mags. The image that we present is 100\% complete down to a magnitude of $I = 20.0$.

Star-galaxy separation is achieved through the size of the object and the use of the {\sc class\_star} parameter within {\sc SExtractor} which, following Pimbblet et al.~(2001), objects greater than 2 arcsec with a {\sc class\_star}  $< 0.1$ are classified as galaxies. Further, we use the {\sc mag\_best} magnitude as our estimate of the total magnitude of a given object.

To include colour information $B$ (sensitive to the young stellar population - late type galaxies) and $R$-band (sensitive to the older stellar population - early type galaxies) photometry from the LARCS catalogue (Pimbblet et al.~2006) was matched to objects in the $I$-band throughout the overlapping field. The colour information was computed by using equi-sized apertures in each filter with a radius equal to 2.5 arcsec, which is worse than the seeing in any of the science images. 

The BCG of A1664 was identified as the brightest spectroscopically confirmed cluster member(Pimbblet et al.~2006). Kirkpatrick et al.~(2009) identified the BCG in A1664 by correlating their optical image with the observed cold gas flow and the co-ordinates of the BCG (13:03:42 -24:14:42) match those in our calibrated $I$-band image (Figure \ref{fig:img}).

\subsubsection{Cluster galaxy identification}
\begin{figure*}
\centering
\includegraphics[scale=1]{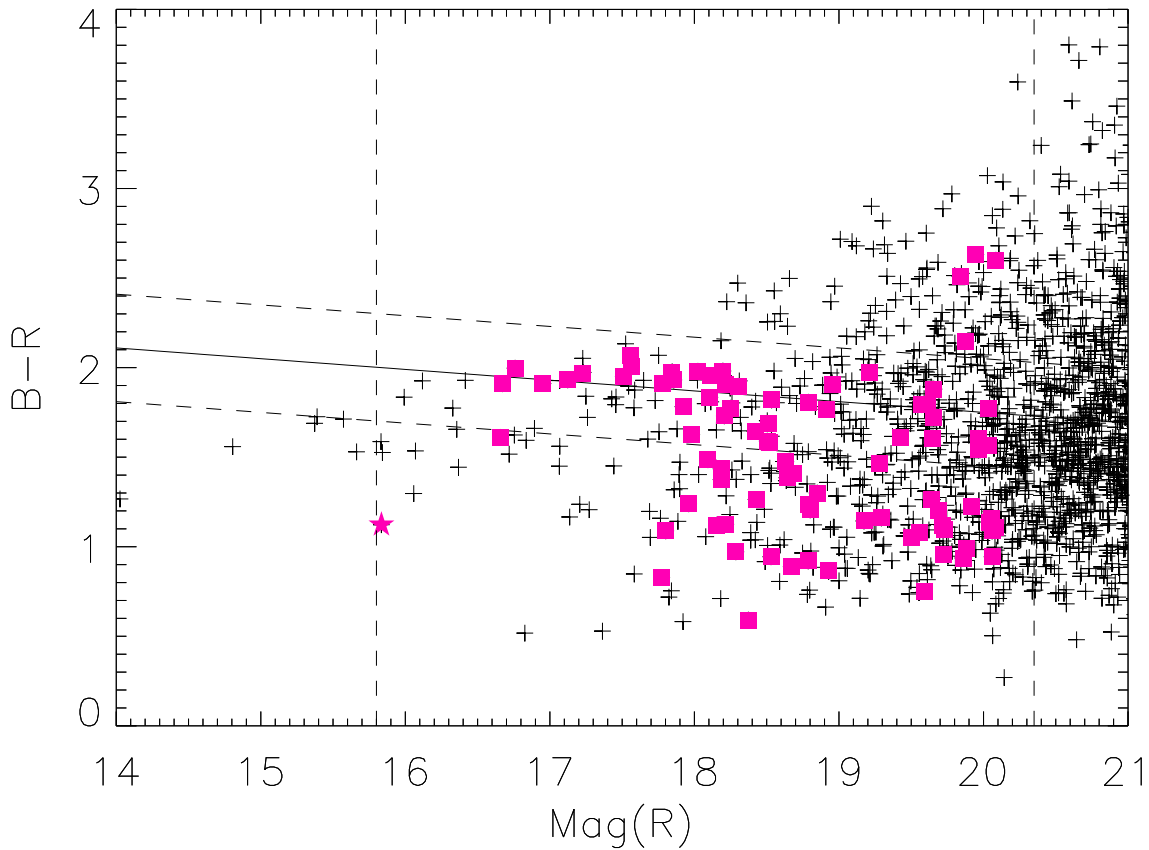}
  \caption{\small{The colour magnitude relation for A1664. The solid line represents the cluster red sequence which has been defined through spectroscopically confirmed-galaxies (over-plotted magenta squares). The dashed parallel lines to the CMR are the 2$\sigma$ limits used to define an envelope for the galaxies to be included in the adaptive smoothing analysis. The BCG is shown with an over-plotted magenta star and galaxies fainter than the BCG down to a magnitude of M* + 1.5 as well as spectroscopically-confirmed blue galaxies have been selected as the input for the adaptive smoothing.}}
  \label{fig:CMR}
\end{figure*}

With a sufficient number of spectroscopically-confirmed cluster galaxies then cluster member identification can be formulated using their redshift, magnitude and colour. However, if sufficient spectroscopic data is lacking, foreground and background interloping galaxies must be eliminated as best as possible through another method in order to minimise contamination in the cluster analysis. We use two different methods using the available spectroscopy and photometry to identify cluster galaxies for two different types of analysis. For the analysis of 2d spatial and luminosity density of the cluster (Section 3.1), we include galaxies from the scatter around red sequence and the available spectroscopy (method (i) below). For the analysis that quantifies the properties of individual galaxies (Sections 3.2), we use a Monte Carlo style statistical correction in colour-magnitude space (method (ii) below).  

i) There exists a well-established correlation between the colour of early-type galaxies and their corresponding luminosity (Sandage \& Visvanathan 1978; Larson et al.~1980). Due to the high number of viralised early-type and S0 galaxies contained in clusters, the majority cluster galaxies will form a tight red sequence relation in colour - magnitude space (known as the colour magnitude relation - CMR). This well defined relation can be used to define which are likely cluster galaxies and which are obvious interlopers. At the same redshift as the cluster, no galaxy can be significantly redder than the cluster's CMR (dust reddening will redden a galaxy by a minimal amount) which means that much redder galaxies must be background objects. Similarly, at the redshift of the cluster all galaxies brighter than the BCG can be excluded as no galaxy which is a part of the cluster can be brighter than the BCG. Figure \ref{fig:CMR} shows the CMR of A1664 and using these relations, we are able to constrain which galaxies are a part of the cluster.

We define galaxies within 2 standard deviations of the scatter of the red sequence down to a luminosity of M* + 1.5 to be cluster galaxies. We also include spectroscopically-confirmed cluster members which are too blue to be included in our red envelope. Some background blue galaxies will be redshifted into our sample but their contribution will be minimal as there are many more cluster galaxies that lie within the limits of the red sequence (Lu et al.~2013). 

ii) Although we have 72 spectroscopically-confirmed galaxies (Pimbblet et al.~2006), there are many galaxies that belong to A1664 that do not have spectroscopic measurements. We use a combination of the spectroscopically-confirmed galaxies and the photometric characteristics to apply a statistical field correction similar to Pimbblet et al.~(2002). Equation \ref{field_corr} assigns the probability \textit{P(C)} of being a cluster member to each galaxy. \textit{N(C + F)} is the number of galaxies in colour-magnitude space known to contain both cluster and field galaxies. \textit{N(F)} is the number of galaxies in colour-magnitude space known to contain purely field galaxies. The field region must be a true representation of a large sample of field galaxies. This is achieved by the selected region containing no cluster galaxies and no other known clusters or dense environments\footnote{Our field region was selected in an analogous method to Pimbblet et al.~(2002) to be 0.131 deg$^{2}$ at a large clustercentric radius.}. The area ratio \textit{A} scales the selected Field region to the same size as the Cluster + Field region. 

\begin{equation}
	P(C)_{col,mag} = \frac{N(C + F)_{col,mag} - N(F)_{col,mag} \times A}{N(C + F)_{col,mag}}
\label{field_corr}	
\end{equation}

The probability of each galaxy is compared to a random number generated between 0 and 1. If \textit{P(C)} $>$ P(random) then the galaxy is retained as a cluster member for the analysis. This process is repeated 100 times in a Monte Carlo fashion to generate the random numbers. As we have limited velocity information available, we denoted the velocity dispersion of the cluster as 1$\sigma$ where any galaxy with a recessional velocity within 3$\sigma$ is assigned a probability of 1, automatically including them as cluster galaxies for every trial. Conversely, galaxies with a recessional velocity outside 3$\sigma$ are assigned a probability of 0 for each trial, ensuring they are never classified as a cluster galaxies. See the appendix in Pimbblet et al.~(2002) for the full method of the statistical field correction without velocity information.

\subsection{Chandra data}
A1664 has been observed twice with {\it Chandra}; a $10\,$ks exposure in June 2001 (ObsID 1648) and a $37\,$ks exposure in December 2006 (ObsID 7901). The {\it Chandra} data were reprocessed using the CHANDRA\_REPRO script within the CIAO software package (version 4.4; Fruscione et al.~2006). The script applies the latest calibrations to the data (CALDB 4.5.1), creates an observation-specific bad pixel file by identifying hot pixels and events associated with cosmic rays (utilising VFAINT observation mode) and filters the event list to include only events with {\it ASCA} grades 0, 2, 3, 4, and 6. The DEFLARE script is then used to identify periods contaminated by background flares and no significant contamination was found. For the imaging analyses, exposure maps which account for the effects of vignetting, quantum efficiency (QE), QE non-uniformity, bad pixels, dithering, and effective area were produced using standard CIAO procedures\footnote{cxc.harvard.edu/ciao/threads/expmap\_acis\_multi/}. The energy dependence of the effective area is accounted for by computing a weighted instrument map with the SHERPA {\sf make\_instmap\_weights} script using an absorbed MEKAL spectral model with $N_{\rm H}=8.95\times 10^{20} {\rm cm}^{-2}$ (Dickey \& Lockman 1990), the average cluster values of $kT = 3.65$\,keV and  abundance 0.46 times solar (Kirkpatrick et al.~2009) and $z=0.128$. Background subtraction for both imaging and spectroscopic analyses was performed using the blank sky backgrounds\footnote{cxc.harvard.edu/contrib/maxim/acisbg/}which were processed in the same manner as the observations. The blank sky backgrounds were reprojected to match the tangent point of the observations, and were normalised to match the $10-12\,$keV counts in the observations. 

\begin{figure*}
\includegraphics[scale=0.5]{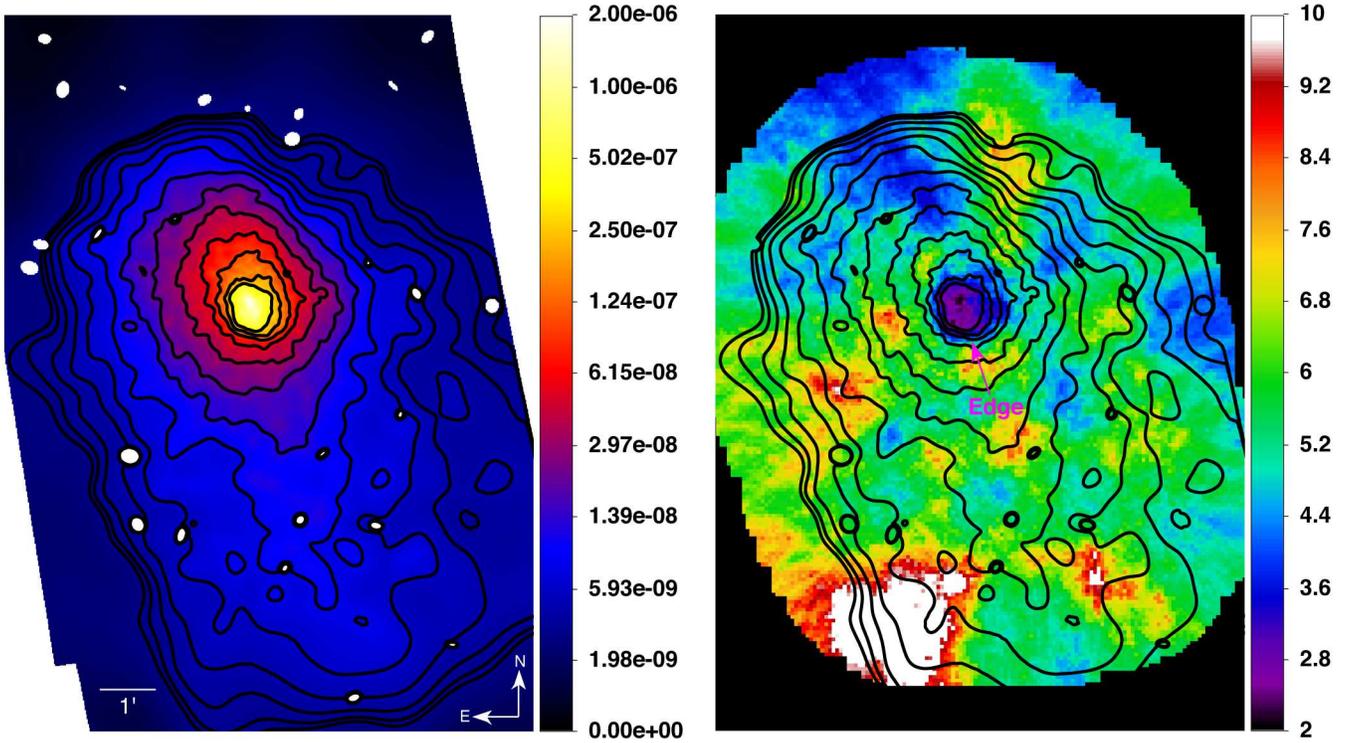}
  \caption{\small{Left is the background subtracted and exposure corrected 0.5-7keV Chandra image. The colour-bar shows the number of photons s$^{-1}$ cm$^{-2}$ and the extension of the ICM emission in the South is clearly visible. On the right is the temperature map where the edge in the cluster core reveal the cold front. The colour-bar is in units of keV and the presence of cold front confirm that A1664 has undergone a recent merger which is past pericentric passage.}}
  \label{fig:xray}
\end{figure*}

Figure \ref{fig:xray} shows the combined, background subtracted and exposure corrected {\it Chandra} image of A1664. The image was adaptively smoothed using the {\sf adaptive\_density\_2d} software within the {\sf TARA}\footnote{http://www2.astro.psu.edu/xray/docs/TARA/} package. The smoothing length was set such that a signal to noise of 10 was achieved for each pixel. Point sources identified with the CIAO tool {\sf wavdetect} were masked prior to smoothing.

A temperature map (Figure \ref{fig:xray}) was created in a similar fashion to Owers et al.~(2013) where a spectra was extracted for each 3.94$\arcsec$ $\times$ 3.94$\arcsec$ pixel using a circular region with radius set such that the total number of background-subtracted counts in the 0.5 -- 7 keV band is $\sim$1000. Spectra are not extracted where the radius required is $\gtsim$ 120$\arcsec$. Responses are extracted on a more coarsely binned ($\sim$16$\arcsec$ $\times$ 16$\arcsec$ pixel) grid. Spectra and responses are extracted for both of the observations. The spectra are then fitted within XPSEC (Version 12.7.1; Arnaud 1996) using an absorbed mekal model (Mewe et al.~1985,1986; Kaastra 1992; Liedahl et al.~1995) with the hydrogen column density fixed at the galactic value (n$_H$ = 8.95 $\times$ 10$^{20}$ cm$^{-2}$), metallicity fixed at 0.45 relative to the solar value (value from Kirkpatrick et al.~2009) and temperature allowed to vary to its best fitting value. Because of the top-hat method of extraction for each pixel, the temperatures are highly correlated where the X-ray brightness is low.

\section{Photometric Analysis}
\subsection{Substructure and sample definition}
To search for substructure, we make use of an adaptive smoothing technique on a conservative selection of cluster galaxies identified through the first method in Section 2.1.1. The adaptive smoothing follows the same two-step procedure outlined in Owers et al.~(2013). Briefly, the spatial distribution of galaxies is first smoothed with a fixed-width Gaussian kernel with an optimal width derived from the data. The resulting smoothed distribution is used to define an adaptive smoothing length at each point which depends on the local galaxy surface density. The resulting smoothed surface density map is sensitive to structure between 200 - 1800 kpc.

\begin{figure}
\includegraphics[scale=0.5]{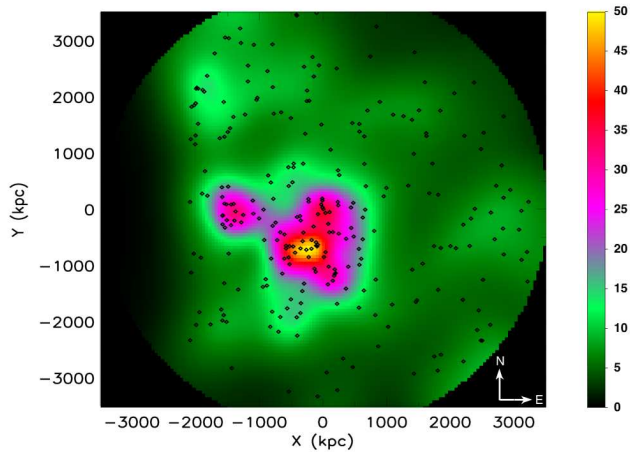}
  \caption{\small{The 2d spatial distribution of adaptively-smoothed cluster galaxies. Galaxies with a B-R colour within 2$\sigma$ of the red sequence and spectroscopically-confirmed blue galaxies down to M* + 1.5 were included for the smoothing. The origin is set to the location of the BCG where the black diamond show the individual galaxies and the colour-bar highlights the number of galaxies per Mpc$^{2}$. The dense feature seen South-West of the cluster core which we identify as best candidate for the merging group.}}
  \label{fig:numdens}
\end{figure}

Figure \ref{fig:numdens} shows the result of the adaptively-smoothed 2d spatial distribution of cluster galaxies centred on the BCG. We identify a local peak in the galaxy number density distribution 800 kpc from the BCG and to the South. The number density in this region is comparable to the cluster core and indicates that there may be substructure present here. Within this, there is an optically bright galaxy (yellow circle in Figure \ref{fig:img}) which is more luminous than all the other galaxies in the substructure. This galaxy has an absolute $R$-band magnitude of -21.39 which is 1.39 mags fainter than the BCG and we consider it to be the brightest group galaxy of this substructure.

\begin{figure}
\includegraphics[scale=0.5]{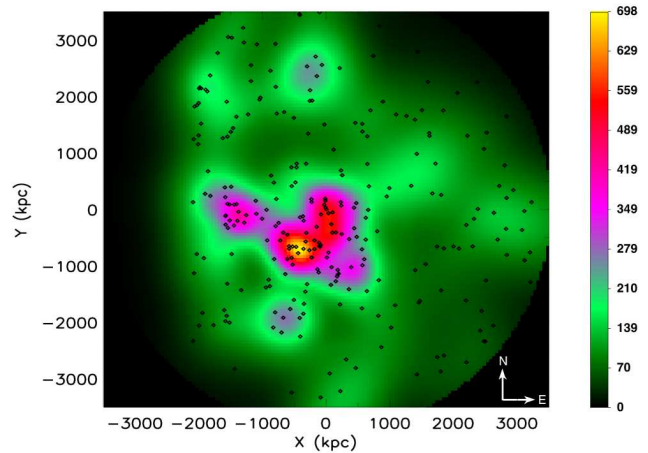}
  \caption{\small{Adaptively-smoothed cluster galaxies where each galaxy has been weighted according to its magnitude relative to the BCG. The same smoothing lengths from Figure \ref{fig:numdens} were used to ensure no artefacts were introduced from the weighting. The symbols are the same as Figure \ref{fig:numdens} where the colour's now show the relative weighted number of galaxies per Mpc$^{2}$.}}
  \label{fig:lumdens}
\end{figure}

A caveat of using the galaxy number density for the adaptive smoothing is that it treats all galaxies as being equal, ignoring the mass of each galaxy. To account for the mass of each galaxy, we assume that galaxy light traces the mass distribution for cluster galaxies. Figure \ref{fig:lumdens} shows the adaptively-smoothed luminosity density for the cluster galaxies in A1664. The same galaxies used in the smoothed galaxy number density (Figure \ref{fig:numdens}) have been weighted according to its luminosity relative to the BCG and smoothed with the same smoothing lengths as the galaxy number density. This method ensures the mass of each galaxy is taken into account and is a more accurate method for searching for substructure. The largest over density  seen in Figure \ref{fig:lumdens} coincides with the largest over density in Figure \ref{fig:numdens} which is strong evidence this feature is the group merging with the cluster. Investigating the properties (i.e. colour and morphology) of the galaxies contained in this group would be an interesting comparison to the cluster galaxies. However, we lack the numbers of spectroscopically confirmed galaxies to separate galaxies belonging to the group from the cluster. Attempting this analysis with the data used in this study would result in group membership being underestimated or contain significant interlopers. Therefore, we leave this task for future work when more spectroscopic data is available.

We relate the position of the detected substructure to the X-ray surface brightness and temperature map (Figure \ref{fig:xray}) to define our galaxy samples. The extended X-ray emission to the South seen in the X-ray surface brightness is likely to be additional emission from the group responsible for the minor merger. If the detected substructure is the group merging with the main cluster then it will be responsible for triggering the cold front shown in the temperature map. With these features in mind, we define two samples to represent different environments within A1664. 

The two samples are shown as blue dashed lines in Figure \ref{fig:img}. The inner sample is the dense regions of A1664 as it includes the cluster core and the detected substructure. The outer sample is the under dense region of A1664 as it is dominated by the cluster outskirts which is sufficiently far away from the cluster core and detected substructure. While maintaining the condition that the inner sample is dense compared to the outer sample, we have run the analysis with different limits for the samples and find no significant change in our results.

\subsection{Galaxy properties}
Here, we investigate how the galaxies have responded to the merger in the two different environments. Cluster membership has been assigned using method ii) in Section 2.1.1 where all uncertainties are 1$\sigma$ errors assigned from the scatter in the Monte Carlo simulation unless stated otherwise. 

\subsubsection{Morphological properties}
One of the most frequently used ways to probe the recent history of galaxies from their photometry
alone is to quantitatively determine their morphology. There are a number of options for doing this, but by far the most widely used and tested is the Concentration-Asymmetry-clumpinesS (CAS; Conselice 2003) approach and its extensions. We search for subtle differences in the $I$-band CAS (Conselice 2003), M$_{20}$ and Gini (Lotz et al.~2004) morphological parameters by measuring the median and skewness of the distribution for each morphological parameter over both samples (Table \ref{tab:morph2}). Abraham et al.~(1996) used a comparable approach (albeit with a slightly different definition of asymmetry and concentration to Conselice 2003) and note that highly skewed asymmetry values are present because their sample is dominated by irregular, peculiar and merging galaxy systems. We also measure the median and skewness of the same morphological parameters of the field galaxies used in section 2.1.1 (Table \ref{tab:morph2}). This provides us with a control sample to contrast our results to where the uncertainty in the field was measured by jackknifing 6 equal areas of the field sample. 

We find that the median of the distribution for all the morphological parameters are within uncertainties between the inner, outer and field samples. The skewness of the distribution for all morphological parameters except asymmetry are also within uncertainties between the samples. This result implies that the population of galaxies is the same in all the galaxy samples down to our spatial resolution limit. The skewness of the asymmetry distribution is significantly higher in the inner sample which indicates there is a greater number of recent, strong galaxy interactions such as galaxy-galaxy mergers occurring. The skewness of the asymmetry distribution in the outer sample resembles the asymmetry distribution of field galaxies which supports the enhancement of asymmetric galaxies found in the inner sample.  

\begin{table*}
\centering
	\caption{\small{The median and skewness of the 5 morphological parameters in the $I$-band in the inner, outer and field sample. The uncertainty in the inner and outer sample were assigned by the 1$\sigma$ scatter from the Monte Carlo simulation and the uncertainty in the field sample was assigned by jackknifing 6 equal areas contained in the field sample from Section 2.1.1. The skewness of the asymmetry is significantly higher in the inner sample than the outer and field samples which can attributed to a higher number of recent strong galaxy interactions. There is no further evidence in the median or skewness in the distribution of the other morphological parameters to suggest the population of galaxies is significantly different between the samples.}} 
	\begin{tabular}{l l l l l l l}
		\hline
		Parameter & Inner Median & Outer Median & Field Median & Inner Skewness & Outer Skewness & Field Skewness\\ [0.5ex]
		\hline
		C & 2.17 $\pm$ 0.03 & 2.21 $\pm$ 0.02 & 2.16 $\pm$ 0.04 & -0.07 $\pm$ 0.08 & 0.39 $\pm$ 0.27 & 0.65 $\pm$ 0.42 \\
		A & 0.20 $\pm$ 0.01 & 0.20 $\pm$ 0.01 & 0.20 $\pm$ 0.01 & 4.43 $\pm$ 0.61 & 1.71 $\pm$ 0.39 & 2.26 $\pm$ 0.79 \\ 
		S & 0.15 $\pm$ 0.01 & 0.14 $\pm$ 0.01 & 0.14 $\pm$ 0.01 & 3.04 $\pm$ 0.53 & -2.45 $\pm$ 1.92 & 2.82 $\pm$ 0.01 \\ 
		M$_{20}$ & -1.71 $\pm$ 0.1 & -1.67 $\pm$ 0.01 & -1.69 $\pm$ 0.01 & 0.96 $\pm$ 0.27 & 2.24 $\pm$ 0.35 & 2.10 $\pm$ 0.21 \\ 
		Gini & 0.52 $\pm$ 0.01 & 0.52 $\pm$ 0.01 & 0.52 $\pm$ 0.01 & 4.19 $\pm$ 0.24 & 5.50 $\pm$ 1.55 & 7.68 $\pm$ 3.93 \\ [1ex]
		\hline
		\end{tabular}
	\label{tab:morph1}
\end{table*}

The CAS system was used by Holwerda et al.~ (2011) to quantify how long neutral hydrogen appears disturbed in galaxy-galaxy mergers. We employ three definitions presented in Holwerda et al.~(2011) to measure the fraction of galaxies that are visibly interacting in our samples (Table \ref{tab:morph2}). Neutral hydrogen is more sensitive to mergers than the stellar population which will result in the neutral hydrogen appearing disturbed for a longer time (Lotz et al.~2010). Therefore, the fraction of interacting galaxies being measured using the stellar population can be treated as a lower limit. Our main finding in Table \ref{tab:morph2} is that the same fraction of galaxies are visibly interacting in both samples. We find the fraction of interacting galaxies varies according to which definition was used to calculate it which will be due to applying the definitions to the optical photometry, coupled with our photometry having a lower spatial resolution than Holwerda et al.~(2011).

\begin{table*}
	\caption{\small{The fractions of galaxies that are visibly interacting according to three definitions presented in Holwerda et al.~ (2011). The same fraction of galaxies are visibly interacting in both samples.}}
	\begin{tabular}{l l l}
		\hline
		Fraction & Inner Sample (\%) & Outer Sample (\%) \\ [0.5ex]
		\hline
		A $>$ 0.4 & 5.3 $\pm$ 1.3 & 8.6 $\pm$ 1.2 \\ 
		G $>$ -0.115M$_{20}$ + 0.384 &16.9 $\pm$ 1.8 & 17.3 $\pm$ 1.6 \\ 
		G $>$ -0.4A + 0.66 & 17.0 $\pm$ 1.8 & 17.25 $\pm$ 1.6 \\ [1ex]
		\hline
		\end{tabular}
	\label{tab:morph2}
\end{table*}

\subsubsection{Galaxy colours}

\begin{figure}
{\includegraphics[scale=0.5]{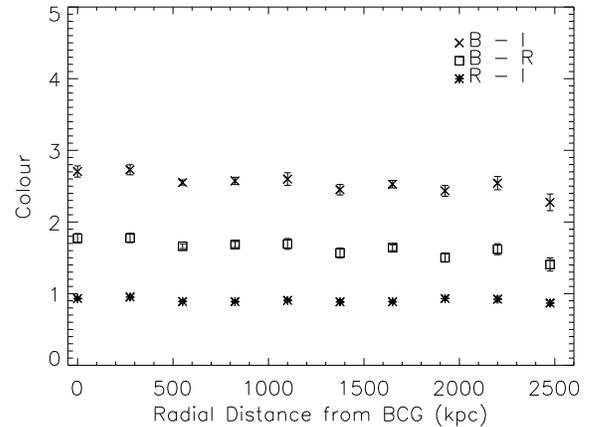}}
\caption{\small{Colour as a function of distance from the BCG where each galaxy was binned into 275 kpc increments. The slight negative trend in the $B$-$I$ colour is consistent with the morphology-density relation and reveals no strong recent star formation at any radii of the cluster}}
  \label{fig:col}
\end{figure}

To probe the effect the cluster merger has on the SFR of the galaxies, we present a three colour ($B$-$I$, $B$-$R$, $R$-$I$) radial analysis in Figure \ref{fig:col}. The galaxies included are contained within the limits of the wedge in Figure \ref{fig:img} and are binned out to a radius of 3 Mpc in 275 kpc bins (similar to Chung et al.~2009). If the merger has a significant effect on the cluster galaxies, this should be reflected by a bluer colour due to starbursting galaxies at the interaction site. There is suggestion of a slight downward trend in the $B$-$I$ colour in figure \ref{fig:col} demonstrating that the galaxies are becoming bluer at large clustercentric radii. This trend is consistent with the morphology-density relation (Dressler 1980) but shows no strong enhancement in recent star formation in any part of the cluster.

We contrast this result to Chung et al.~(2009), who investigated galaxy colour as a function of radial distance in the Bullet cluster. They found that a shock propagating through the ICM was not powerful enough to induce a starburst from the rapid increase in pressure on the galaxies, resulting in no significant colour change. This demonstrates that galaxy colour can be uniform in different environments within the Bullet cluster which is a more hostile environment than A1664.

\begin{figure}
\centering
{\includegraphics[scale=0.5]{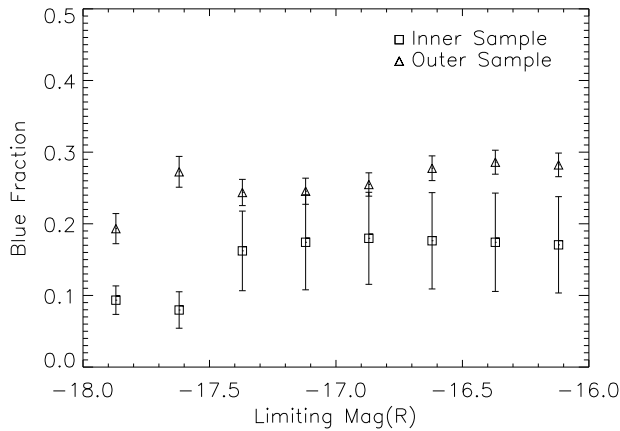}}
\caption{\small{The blue fraction for the two samples as a function of limiting magnitude. The outer sample has significantly higher blue fraction at bright magnitudes and continues to have a higher blue fraction (though not as significant) at faint magnitudes.}}
  \label{fig:bf}
\end{figure}

We have calculated the Butcher-Oemler blue fraction which defines a galaxy as blue if it is 0.2 bluer than the cluster red sequence in the $B$-$V$ rest frame (Butcher \& Oemler 1978, 1984). We present Figure \ref{fig:bf} which plots the blue fraction as a function of limiting magnitude and find there is consistently a higher blue fraction in the outer sample down to all limiting magnitudes. This trend is consistent with the radial colour analysis (Figure \ref{fig:col}) emphasising that bluer galaxies preferentially reside in less dense environments such as the outskirts of clusters.

\section{Discussion}
There are three pieces of evidence in favour of interpreting A1664 as presently undergoing a minor merger. These are: (i) the extended feature to the South in the X-ray surface brightness image (Figure \ref{fig:xray}), (ii) the cold front in the cluster core seen in the temperature map and (iii) the detected substructure displayed in the adaptively-smoothed unweighted and weighted 2d spatial distribution of cluster galaxies (Figure \ref{fig:numdens} and \ref{fig:lumdens}). In this section, we discuss the limitations of our data in conjunction with the effects a minor merger has on cluster galaxies.

If the substructure to the South is the perturber, then it must have passed pericentre as it would be responsible for triggering the cold front in the cluster core. As many cluster galaxies lack redshifts, we are unable to unambiguously confirm that the substructure in Figures \ref{fig:numdens} and \ref{fig:lumdens} is the remnant peturber. However this is the most likely scenario for the following reasons: i) The X-ray morphology which appears extended towards the BCG is best explained by  emission from the intra group medium of the peturber. ii) The bright galaxy in the yellow circle of Figure \ref{fig:img} appears luminous enough to be the brightest group galaxy. 

We measure the skewness of the CAS (Conselice 2003), M$_{20}$ and Gini (Lotz et al.~2004) parameters. This method will be most sensitive to the differences in the galaxy population between the samples as well as galaxies with measured morphological values in the extremities of the distribution. We interpret the higher percentage of asymmetric galaxies found in the inner sample as a result of increased galaxy-galaxy interactions while the group passes through pericentre. As the merging group has travelled passed pericentre, this interpretation is consistent with the work of Vijayaraghavan \& Ricker (2013) who showed that galaxy interactions increased during pericentric passage. Due to a higher percentage of late type galaxies residing in the field than clusters, we would expect the field to have a higher skewness of the asymmetric parameter than a relaxed cluster. However,  A1664 is not a relaxed cluster and we find a higher skewness of asymmetry in the inner sample which supports the previous result that a cluster mechanism (galaxy-galaxy mergers) is responsible for this increase in asymmetric galaxies. 

The major limitation of this morphological analysis is the spatial resolution of our $I$-band photometry. The biggest consequence is that only the galaxies currently undergoing strong interactions (i.e. recent mergers) will be differentiated from the cluster population. This limitation is the most likely explanation for detecting a difference between the galaxy populations in the asymmetry parameter but not the in the other morphological parameters (Table \ref{tab:morph1}) and the disagreement between the definitions of interacting galaxies in Table \ref{tab:morph2}. 

If galaxies in the cluster were undergoing a starburst, we would expect an enhancement in the blue section of Figure \ref{fig:col}. We observe a slight negative trend in the most sensitive colour ($B$-$I$) but see no significant enhancements of recent star formation in any part of the cluster. We expect to find a similar radial colour trend in relaxed clusters which implies other galaxy transformation mechanisms (such as pre-processing) are able to quench star formation well before interacting with the merging group.

We consistently measure a higher blue fraction in the outer sample of A1664 (Figure \ref{fig:bf}) where the clearest difference is seen in the bright magnitudes. The morphology-density relation (Dressler 1980) predicts the trend we see in the measured blue fraction. As we measure the blue fraction down to fainter magnitudes, we include less luminous (and therefore less massive) galaxies  in our sample which are more sensitive to interactions. Due to this sensitivity, starbursts could be more easily triggered in less massive galaxies. However, the fainter we measure our blue fraction to, the less complete our sample is which increases the uncertainty in our measurements. The colour analysis has shown that A1664 has the same colour properties as a cluster that is not undergoing a minor merger.

\section{Conclusions}
We have presented a detailed analysis using multi-band photometry of the disturbed galaxy cluster A1664. Our main findings from this study are:

\begin{itemize}
\item We detect substructure in the smoothed unweighted and weighted 2d galaxy distributions. The substructure coincides with the extended X-ray emission in the South and a local over-density in the optical image. These results are best explained by the substructure being the remnant core of the merging group where the bright optical galaxy is most likely the surviving BCG. 
\item We find there is a higher percentage of morphologically asymmetric galaxies in the inner region of A1664 when compared to the outer and field samples. The difference between the inner and outer sample is significant to 3.7$\sigma$. Due to the limited resolution of our imagery, our morphological analysis will only be able to detect galaxies that are highly asymmetric. It is likely that these galaxies have undergone strong encounters resulting from a higher interaction rate as the merging group passes through pericentre.
\item We only detect a variation in the asymmetry parameter between our samples. We believe our spatial resolution was the main reason for not being able to detect variation in the other morphological parameters. 
\item There is disagreement between the percentage of galaxies interacting according to which definition is implemented. This is also most likely due to the spatial resolution of our photometry.
\item The radial colour profile of A1664 has a slight negative trend in the $B$-$I$ colour and a flat trend in the $B$-$R$ and the $R$-$I$ colours. This colour profile shows no enhancement at a specific radial distance from the BCG which indicates no significant recent star formation occurring in the cluster. 
\item The blue fraction was found to be higher in the outer sample which is consistent with blue galaxies preferentially residing in the outskirts of clusters.
\end{itemize}

Our work shows that a minor merger has subtle effects on the population of cluster galaxies in A1664 -- in the sense that only one interaction signature of many (Holwerda et al.~2011) demonstrated any significant consequence for the galaxies involved. Our morphological analysis is found to be consistent with the work of Vijayaraghavan \& Ricker (2013) who show that galaxy-galaxy interactions increase while a group passes through pericentre. The colour of the galaxies in A1664 resembles what we would expect to find in a relaxed (i.e. not merging) cluster. As the population of cluster galaxies is found to be quite uniform, we have not observed any significant galaxy transformations in this environment. It is likely that galaxies are being pre-processed well before they are able to interact with the merging group but it is possible that cluster mechanisms such as rapid strangulation or ram pressure stripping are contributing to the uniformity of cluster galaxies. The techniques used in this analysis can be applied to other clusters with signatures of a minor merger (such as cold fronts). The application to other clusters could help disentangle the effect minor mergers have on clusters which are difficult to observationally detect.

\section*{Acknowledgments}
This work is based on observations made with WFI using the 40$^{\prime\prime}$ telescope at Siding Spring Observatory. We extend our warm thanks to the MSSSO Research School of Astronomy and Astrophysics Time Allocation Committee. Our thanks also go to John Shobbrook for sharing his expertise and training us to use WFI on the 40$^{\prime\prime}$ telescope.

This research has made use of the USNOFS Image and Catalogue Archive operated by the United States Naval Observatory, Flagstaff Station (http://www.nofs.navy.mil/data/fchpix/). The scientific results reported in this article has made use of software provided by the Chandra X-ray Center (CXC) in the application packages CIAO, ChIPS, and Sherpa and also of data obtained from the Chandra archive at the NASA Chandra X-ray Center (cxc.harvard.edu/cda/).

We thank the referee for their insightful comments that has helped improve the original manuscript. D.K. acknowledges support from an Australian postgraduate award (APA). KAP thanks Christ Church College, Oxford, for their hospitality whilst some of this work was being undertaken and is grateful to Oxford university for support during a sabbatical visit. M.S.O  acknowledges the funding support from the Australian Research Council through a Super Science Fellowship (ARC FS110200023).

\end{document}